\documentclass[letterpaper,4pt,times,preprint]{elsarticle}

\journal{Acta Materialia}
    
    \usepackage{tgtermes}
    \usepackage{tikz}
    \usepackage{hyperref}
    \usepackage[version=3]{mhchem}

    \usepackage{siunitx}
    \usepackage[letterpaper, portrait, margin=0.60in]{geometry}

    \usepackage{hhline}  
    \usepackage{float}
    \usepackage{multirow}
    \usepackage{graphicx,booktabs}
    \usepackage{booktabs,tabularx}
    \usepackage{xfrac}
    \usepackage{titlesec}
    \usepackage{wrapfig,booktabs}
    \usepackage{mwe}
    \usepackage{adjustbox}
    \usepackage{caption}
    \usepackage{subcaption}
    \usepackage{makecell}
    \usepackage{rotating}
    \usepackage[permil]{overpic}
    \usepackage{pict2e} 

    \usepackage{amsmath,array, amssymb}
    \DeclareMathOperator*{\argmin}{\textbf{arg}\,\textbf{min}}
    \usepackage{xcolor,soul}
    \usepackage{tcolorbox}
    \usepackage{calrsfs}
    \usepackage{longtable}
    \usepackage{ragged2e}
    \usepackage[normalem]{ulem}
    
\hypersetup{colorlinks,breaklinks,
            urlcolor=Maroon,
            linkcolor=Maroon}
    
    \def\correction#1{%
    \abovedisplayshortskip=#1\baselineskip\relax\belowdisplayshortskip=#1\baselineskip\relax%
    \abovedisplayskip=#1\baselineskip\relax\belowdisplayskip=#1\baselineskip\relax}

    \fboxsep=\tabcolsep\relax
    \fboxrule=\arrayrulewidth\relax

    \newcolumntype{A}[2]{%
    >{\minipage{\dimexpr#1\linewidth-2\tabcolsep-#2\arrayrulewidth\relax}\vspace\tabcolsep}%
    c<{\vspace\tabcolsep\endminipage}}

    \usepackage{siunitx}

    \usepackage{tikz}
    \usepackage{color}
    \definecolor{scalebgcolor}{rgb}{0.08,0.52,0.80}     
\biboptions{numbers,sort&compress}

\begin{document}

\begin{frontmatter}
	
\title{Compositionally Grading Alloy Stacking Fault Energy using Autonomous Path Planning and Additive Manufacturing with Elemental Powders}

\author{James Hanagan$^{a}$, Nicole Person$^{a}$, Daniel Salas$^{a}$, Marshall Allen$^{a,b}$, Wenle Xu$^{a}$, Daniel Lewis$^{a}$, Brady Butler$^{a,c}$, James D. Paramore$^{a,d}$, George Pharr$^{a}$, Ibrahim Karaman$^{a}$, Raymundo Arr\'{o}yave$^{a}$}
\address{$^a$ Department of Materials Science and Engineering, Texas A\&M University, College Station, TX 77843, USA}
\address{$^b$ J. Mike Walker '66 Department of Mechanical Engineering, Texas A\&M University, College Station, TX 77843, USA}
\address{$^c$ DEVCOM Army Research Laboratory, ARL South, College Station, TX 77843, USA}
\address{$^d$ George H.W. Bush Combat Development Complex (BCDC) at the Texas A\&M RELLIS Campus, Bryan, TX 77807 , USA}

\cortext[mycorrespondingauthor]{Corresponding author email:\textrm{rarroyave@tamu.edu}} 

\begin{abstract}

Compositionally graded alloys (CGAs) are often proposed for use in structural components where the combination of two or more alloys within a single part can yield substantial enhancement in performance and functionality. For these applications, numerous design methodologies have been developed, one of the most sophisticated being the application of path planning algorithms originally designed for robotics to solve CGA design problems. In addition to the traditional application to structural components, this work proposes and demonstrates the application of this CGA design framework to rapid alloy design, synthesis, and characterization. A composition gradient in the CoCrFeNi alloy space was planned between the maximum and minimum stacking fault energy (SFE) as predicted by a previously developed model in a face-centered cubic (FCC) high entropy alloy (HEA) space. The path was designed to be monotonic in SFE and avoid regions that did not meet FCC phase fraction and solidification range constraints predicted by CALculation of PHase Diagrams (CALPHAD). Compositions from the path were selected to produce a linear gradient in SFE, and the CGA was built using laser directed energy deposition (L-DED). The resulting gradient was characterized for microstructure and mechanical properties, including hardness, elastic modulus, and strain rate sensitivity. Despite being predicted to contain a single FCC phase throughout the gradient, part of the CGA underwent a martensitic transformation, thereby demonstrating a limitation of using equilibrium CALPHAD calculations for phase stability predictions. More broadly, this demonstrates the ability of the methods employed to bring attention to blind spots in alloy models.

\end{abstract}

    
\end{frontmatter}

\section{Introduction}
Emphasis in alloy design has been shifting for some years now from trial and error synthesis and experiments to relying more heavily on models to screen and target compositions for synthesis and experimentation. Experiments are still the ultimate test for the viability of any given composition, but alloy property models and design indicators allow for narrowing from hundreds of thousands of possible compositions down to a reasonable number for testing. What defines the quantity of compositions deemed ``reasonable'' for testing is heavily dependent on the rate at which compositions can be synthesized, tested, and characterized. Compositional gradients are an excellent candidate for rapid synthesis and testing of a series of related compositions. They allow for not only the rapid synthesis of composition after composition, but also for the characterization of each composition in quick succession with techniques such as microscopy, X-ray diffraction, microhardness, and nanoindentation. Other strategies for high-throughput, bulk alloy synthesis have been tested, including building a series of discrete alloy compositions directly on substrates with laser directed energy deposition (L-DED) \cite{vecchio2021high, moorehead2020high}, hot isostatic pressing (HIP) of various compositions into a single sample \cite{zhao2021high}, and vacuum arc melting of multiple alloy compositions at once in the form of design campaigns \cite{acemi2024multi, broucek2024design, mulukutla2024illustrating, vela2023data, vela2023high}. Alloy composition gradients offer an advantage over existing methods of discrete composition libraries, which is the ability to test specific property models by designing a gradient in the chosen property across the length of the sample and measuring the behavior as a function of position and, therefore composition. This is the motivation for this work where the property model of interest is stacking fault energy (SFE).

SFE is an important material property that has great potential for use in alloy design frameworks because of its ability to describe how an alloy is likely to accommodate plastic deformation and its implications on alloy strengthening and ductility \cite{shih2021stacking, kivy2017generalized, grassel2000high, zhao2006tailoring}. Mechanisms of plastic deformation in face-centered cubic (FCC) alloys include dislocation glide, twinning-induced plasticity (TWIP), and transformation-induced plasticity (TRIP), each of which has been shown to correlate to different ranges of SFE \cite{de2018twinning, lee2010correlation, karaman2001competing, karaman2000deformation, karaman2000modeling, grassel1998effect, picak2023orientation, khan2022towards, kivy2017generalized, zhao2006tailoring}. A model for predicting SFE was previously developed as a support vector regression (SVR) on a series of selected features of alloys in the FCC CoCrFeMnNiV-Al alloy space \cite{khan2022towards}. In order to determine where this model's limitations exist for predicting SFE, a series of compositions from the model with a wide range of SFE values must be tested. To examine as many compositions as possible from this model, a method for synthesizing and characterizing compositions in a high-throughput manner is needed. For this, a linear gradient in model-predicted SFE is proposed. The composition gradient can be built rapidly using multi-material L-DED, allowing each composition to be characterized for both microstructure and selected mechanical properties in quick succession.

Interest in compositionally graded alloys (CGAs) as a subset of the broader category of functionally graded materials (FGMs) has grown substantially over the past decade due to the need for components that can withstand demanding environments and overcome property trade-offs arising from using a single alloy for a given application. As a result, a number of design methodologies have come about for these types of gradients. The simplest of these is linearly grading composition between two selected endpoints, often conventional alloys such as Invar 36, 304L stainless steel, Ti-6Al-4V, and Inconel 625 \cite{bobbio2017additive, carroll2016functionally}. A key drawback to this method is that gradients often fail at intermediate compositions, usually from the formation of deleterious phases that are susceptible to cracking \cite{bobbio2017additive, meng2019additive}. Another method utilizes phase diagrams for mapping gradients, typically by selecting intermediate compositions to linearly grade between and avoid the formation of deleterious phases \cite{hofmann2014compositionally, bobbio2022design}. A disadvantage to this method is that it requires the gradient to be planned in an alloy design space that can easily be visualized in two dimensions. This requires either confining oneself to a single ternary alloy system \cite{hofmann2014compositionally}, or creatively arranging multiple diagrams together and planning the gradient between those \cite{bobbio2022design}. 

To resolve the issues above, gradients can be designed utilizing path planning algorithms that can take into account defined constraints and objectives of the compositional gradient and optimize properties of the path itself \cite{kirkComputationalDesignGradient2018, kirkComputationalDesignCompositionally2020d, eliseeva2019functionally}. This method is advantageous because it can be expanded to higher dimensional alloy systems and intelligently design the intermediate compositions to not only avoid undesired compositions and phases, but also optimize path properties such as path length and monotonicity of a given material property \cite{kirkComputationalDesignGradient2018, kirkComputationalDesignCompositionally2020d}. This is the method employed in the present work, where a monotonic gradient between the minimum and maximum SFE in a constrained CoCrFeNi alloy space has been planned using a form of the rapidly-exploring random trees (RRT) algorithm \cite{karamanSamplingbasedAlgorithmsOptimal2011a}, and a linear gradient in SFE has been synthesized and characterized.

\section{Methods and Materials}
\subsection{{Computational Methods}}
The gradient design methodology used in this work was adapted from \cite{kirkComputationalDesignGradient2018}. Following this methodology, the fixed-node implementation of the probabilistically optimal rapidly-exploring random trees algorithm (RRT*FN) \cite{karamanSamplingbasedAlgorithmsOptimal2011a, adiyatovRapidlyexploringRandomTree2013} was applied in the CoCrFeNi alloy design space using a subspace-inclusive sampling scheme from \cite{allenSubspaceInclusiveSamplingMethod2022b}. With this approach, the alloy composition state space $Z$ was divided into subsets $Z_\text{obs}$ and $Z_\text{free}$, where $Z_\text{free} = Z \setminus Z_\text{obs}$ and $Z_\text{obs}$ defined the obstacle region. That is, $Z_\text{obs}$ consisted of all compositions that did not meet constraints for phase makeup and solidification range. More specifically, the constraints were that all feasible compositions in the design space must be predicted to have a single FCC phase stable at 600~\textdegree{C}, and a solidification range ($T_\text{liquidus} - T_\text{solidus}$) less than 50~\textdegree{C} as predicted by CALculation of PHAse Diagrams (CALPHAD). The phase constraint was enforced by requiring $>$ 0.99 mole fraction of predicted FCC phase, $f_\text{FCC}$, at 600~\textdegree{C}. Inversely, the subset of compositions that satisfy these constraints was given by $Z_\text{free}$. A hypothetical compositional gradient was represented by a continuous path $\sigma$ in the design space $Z$. This path $\sigma$ is a continuous function $\sigma:[0,1]\rightarrow z$ that relates a normalized path index $\alpha \in [0,1]$ to a point $z$, or alloy composition, in the state space $Z$. The objective of the FGM design problem was to minimize the user-defined cost function $c(\sigma)$ by determining the optimal gradient path, $\sigma_\text{best}$, between two prescribed compositional endpoints, $z_\text{init}$ and $z_\text{goal}$. Thus, the overall design problem was defined as follows:

\begin{align*}
& \textbf{Find}
& & \sigma_\text{best} = \argmin_{\sigma} \; c(\sigma) \\
& \textbf{Subject to}
& & \sigma(\alpha) \in Z_\text{free} \; \forall \, \alpha \in [0,1], \\
&&& Z_\text{free} = \Big\{z: f_\text{FCC}(z)>0.99 \; , (T_{L} - T_{S}) < 50^\circ\text{C} \Big\},\\
&&& \sigma(0) = z_\text{init} = z_\text{min SFE}, \\
&&& \sigma(1) = z_\text{goal} = z_\text{max SFE}.
\end{align*}

\noindent The desired endpoints ($z_\text{init}$ and $z_\text{goal}$) for this problem were the compositions with the maximum and minimum SFE that met the constraints on the design space. These endpoints were chosen to capture the greatest range of values from the SFE model and determine where the model's limits for describing the actual behavior of alloys existed. A lack of monotonicity (LOM) cost function \cite{kirkComputationalDesignCompositionally2020d} was defined to enforce a path where the SFE value is always increasing by penalizing any lack of increase (LOI) in the predicted SFE value along the path, since the starting composition was the composition with the lowest SFE. LOI is simply the negative part of the first derivative of SFE with respect to the path index, $\alpha$, integrated over the path. A monotonic property path allows for one-to-one matching of SFE values to compositions, which can be easily translated to a linear property gradient for L-DED by selecting compositions at the desired intervals of SFE. This cost function formulation also included a path length term, $l$, weighted by $w$, so the algorithm would minimize the length of the path in composition space once a monotonic SFE path was found. Minimizing path length effectively minimizes the number of layers required to print the gradient by reducing unnecessary compositional complexity in the path. This cost function was formulated as:

\begin{equation}
c(\sigma) = \mathrm{LOI_{\sigma}}(p) + wl = \int_{0}^{1} \left( \frac{\text{d}p}{\text{d}\alpha}\right)^{-} \, \mathrm{d}\alpha + wl
\end{equation}

In this work, the value of $w$ was set to $10^{-6}$ based on the parameter study conducted in \cite{kirkComputationalDesignCompositionally2020d}. This value ensured that the magnitude of the length term was far too small to offset the primary objective of finding a gradient monotonic in SFE.

\begin{figure}
    \centering
    \includegraphics[width=1\textwidth]{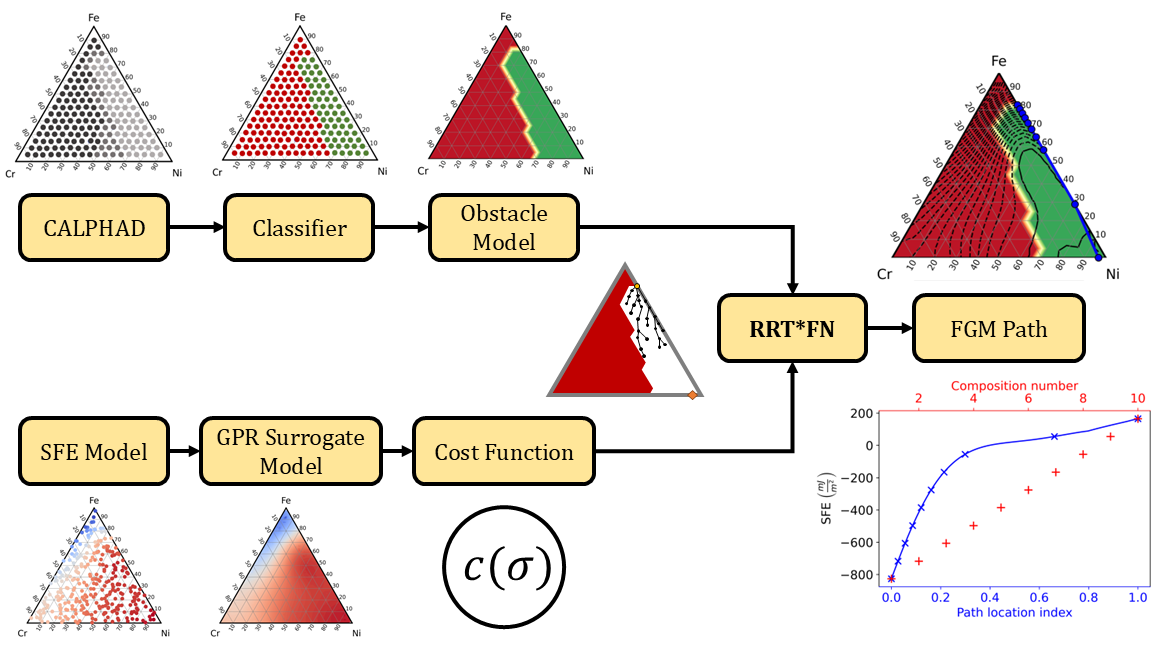}
    \caption{Flow diagram of the computational FGM design methodology adapted from \cite{kirkComputationalDesignGradient2018, kirkComputationalDesignCompositionally2020d, allenSubspaceInclusiveSamplingMethod2022b} used in this work. CALPHAD phase equilibrium and solidification data is used to train machine learning classifiers for the alloy system and all subspaces. These classifiers are used as the obstacle model for the robotic path planning algorithm. The SFE model from \cite{khan2022towards} is used to build a computationally efficient surrogate model used for calculating the LOI term in the cost function. The cost function and obstacle models are sampled by the RRT*FN path planning algorithm to find the optimal FGM path through the alloy composition space.}
    \label{fig:algorithm}
\end{figure}

The design algorithm, visualized in Fig.~\ref{fig:algorithm}, uses the following inputs to search for $\sigma_\text{best}$: binary obstacle classification models for each subspace, two compositional endpoints for the gradient path ($z_\text{init}$ and $z_\text{goal}$), and the cost function $c(\sigma)$. The path planning algorithm, RRT*FN, is a sampling-based method that sequentially adds samples to an optimized tree of nodes/compositions and edges between nodes until it has reached a criterion for completion. RRT*FN operates by sampling a new random point $z_\text{new}$, in this case an alloy composition. The algorithm then finds the nearest existing node in the tree $z_\text{nearest}$. Based on a prescribed maximum step size $r_{\text{step}}$, the node is moved or ``steered'' closer to $z_\text{nearest}$ (at a distance of $r_{\text{step}}$) if it is too far away. Based on a prescribed neighborhood distance $r_{\text{neighborhood}}$, the potential edge between the new node and each node in the neighborhood is checked against the obstacle model to ensure it does not traverse $Z_\text{obs}$. The cost is then calculated from the starting endpoint node $z_\text{init}$ through the tree to the current node when hypothetically connected to each feasible neighbor. The feasible neighboring node with the lowest cost becomes the parent node and an edge is added between the parent node and $z_\text{new}$. Then, the tree is ``rewired'' by checking the cost from $z_\text{init}$ to the neighbors of $z_\text{new}$. If the cost to reach any of the neighboring nodes is reduced with $z_\text{new}$ as the parent, and such an edge does not traverse $Z_\text{obs}$, the tree is ''rewired'' by creating an edge between the neighbor and $z_\text{new}$, and the neighbor's existing edge to its parent is deleted. Lastly, the fixed-node implementation includes an additional global node removal step, where once a specified maximum number of nodes are reached, one node that is not a parent is randomly selected and deleted to retain the same amount of nodes for memory efficiency. This process describes all steps in the algorithm, and continues for a specified number of iterations or until meeting the completion criterion. Further discussion and illustrations of this algorithm can be found in the original work by Adiyatov and Varol \cite{adiyatovRapidlyexploringRandomTree2013}.

In this work, CALPHAD predictions for phase stability and solidification range were done using high-throughput property and equilibrium calculation scripts with Thermo-Calc's TC-Python API and TCHEA4 database \cite{ThermoCalcSoftwareHigh, maoTCHEA1ThermodynamicDatabase2017}. Over 73,000 compositions in the CoCrFeNi space were generated using Halton sequence sampling, a pseudo-random space-filling sampling technique. These compositions were then assessed via Thermo-Calc to find all relevant values for determining the obstacle region. This composition data was then labeled based on the CALPHAD results and the constraints to separate the obstacle and free points. This labeled data was inputted into subspace-specific $K$-nearest neighbors classifiers (with parameter $K=3$) to create the binary obstacle classification models for the CoCrFeNi alloy space and all subspaces respectively. The SFE model is an SVR that was fit on a set of alloy properties from density functional theory (DFT) calculations for 500 alloys in the FCC CoCrFeMnNiV-Al alloy space \cite{khan2022towards}. SFEs for each alloy were calculated with DFT as the difference in energy between FCC stacking and hexagonal closed-packed (HCP) or double hexagonal close-packed (DHCP) stacking. For more detail on the SFE model, refer to \cite{khan2022towards}. Once the obstacle region for the space was defined, the compositions with the minimum and maximum SFE in the free design space could be determined. The SVR model was used to calculate the SFE for all feasible CALPHAD compositions. Only CALPHAD compositions were considered as end-points since they had been specifically assessed based on the phase and solidification constraints. The feasible compositions with the extreme SFE values were found to be Fe-17.3~at.\%~Ni with a predicted SFE of -826.2~$\mathrm{{mJ}/{m^2}}$ and Ni-3.4~at.\%~Cr with a predicted SFE of 165.4~$\mathrm{{mJ}/{m^2}}$. Due to the high number of model queries required by RRT*FN to repeatedly calculate the cost function for new edges, the query time of the SVR became prohibitive. Thus, it was necessary to train a faster surrogate model to ensure the RRT*FN algorithm could run for a sufficient number of iterations. To train this surrogate, the SFE model was queried at random for a set of 14,605 compositions. These were then used to train a Gaussian process regressor (GPR) to create a continuous model to be sampled by the path planning algorithm. 

\begin{figure}
\begin{subfigure}{.5\textwidth}
  \centering
  \includegraphics[width=\linewidth]{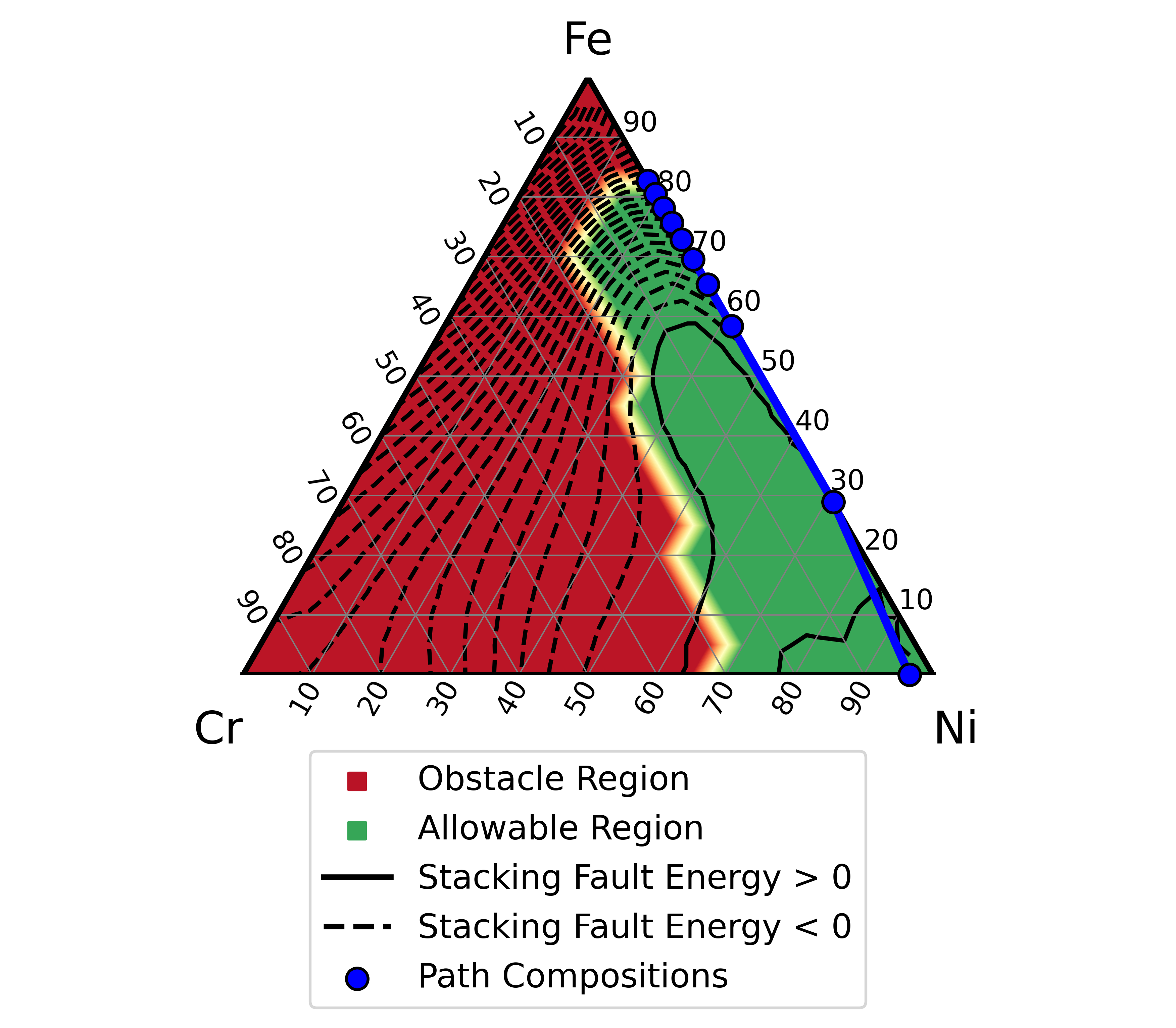}
  \caption{Final path in the CrFeNi ternary}
  \label{fig:sfigTern}
\end{subfigure}%
\begin{subfigure}{.5\textwidth}
  \centering
  \includegraphics[width=\linewidth]{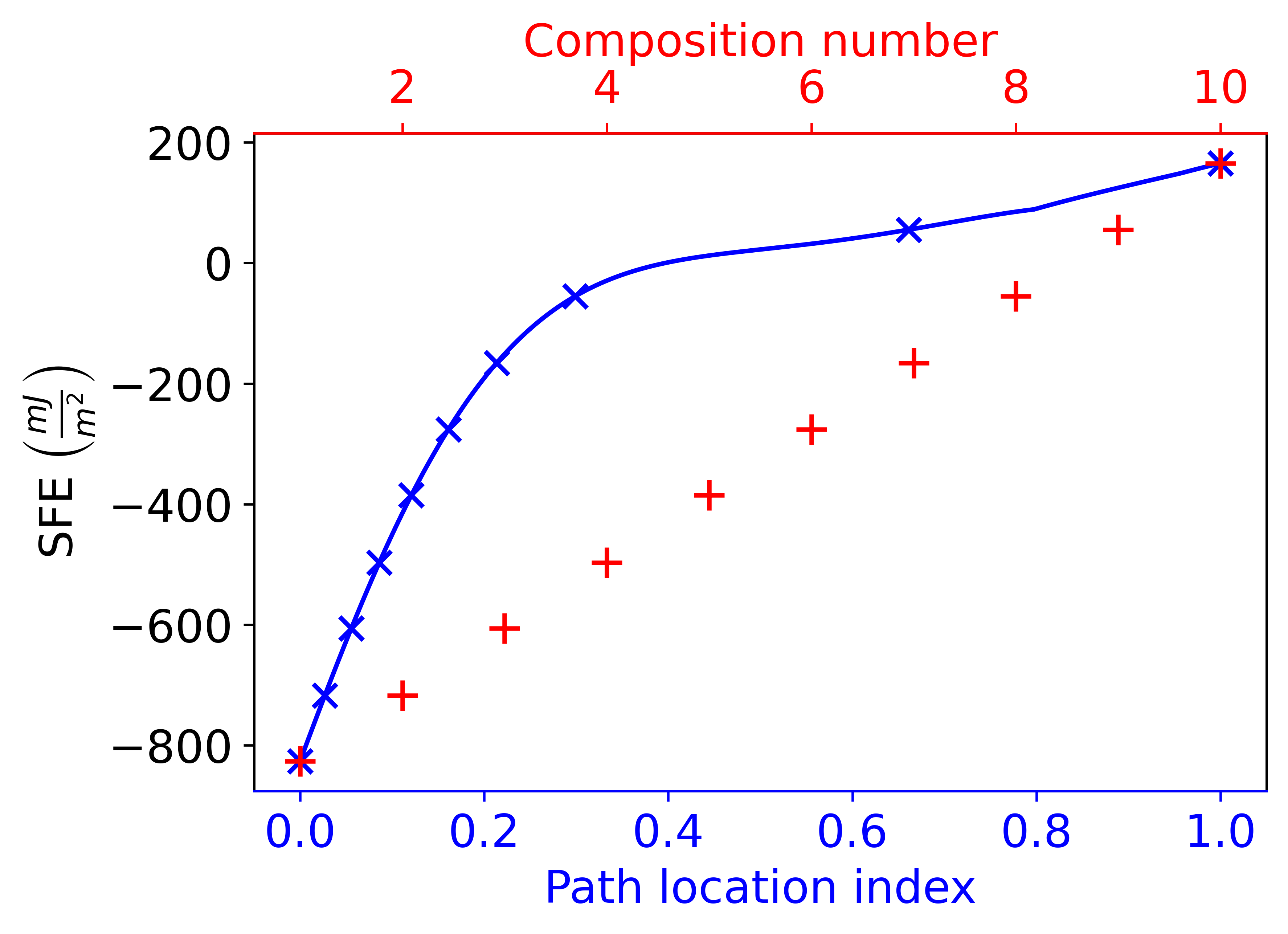}
  \caption{Sampling the monotonic path to obtain a linear SFE gradient}
  \label{fig:sfigLin}
\end{subfigure}
\caption{Plots of the final path as planned by the RRT*FN algorithm (a) in the CrFeNi ternary space and (b) sampling from the path to yield a linear gradient in SFE.}
\label{fig:finalpath}
\end{figure}

The RRT*FN algorithm was run for a total of 37,000 iterations with the inputs above. The resulting path is plotted in Fig. \ref{fig:finalpath}, with exact compositions listed in Table \ref{Table:AtomicComposition}. Despite the path being planned in the quaternary CoCrFeNi space, the final path only contained compositions in the ternary CrFeNi space. Because of this, the path can be easily visualized on a CrFeNi ternary diagram in Fig. \ref{fig:sfigTern}. The path likely excluded any compositions with Co because neither endpoint contained Co, and the final path could easily achieve the defined objectives without this extra degree of freedom. After obtaining the monotonic SFE path in composition space, it was sub-sampled into ten compositions with a linear trend in SFE to be built in physical space. This involved dividing the overall difference in SFE across the gradient by the appropriate number of segments and finding the composition that corresponds to each SFE value. This is plotted in Fig. \ref{fig:sfigLin}. The composition index numbers each of the ten compositions selected for the final gradient, while the path location index refers to each composition's location on the path. This index is on a scale from 0 to 1 and denotes the proximity to either endpoint in composition space. The exact values of each composition are listed in Table \ref{Table:AtomicComposition}. It can be observed that eight of the ten compositions contained more than 50~at.\% Fe. This is because the most rapid increase in SFE along the path in composition space occurs in the majority Fe compositions.

\subsection{{Sample Fabrication}}

The samples were manufactured via L-DED with a combination of high-purity Fe, Ni and Cr elemental powders in an Optomec LENS MR-7 L-DED system. These high-purity elemental powders ($>$99 wt.\% purity) with particle sizes ranging from 45 to 106~\unit{\um} were acquired from ECKART GmbH. Note that the LENS system has 4 different powder hoppers and nozzles that can be independently configured to, either individually or simultaneously, feed powder at various mass flow rates by controlling the rotation speed of the hoppers (RPM). 

The first version of the gradient was divided into 10 sections with a different target composition for each section, based on the calculated optimal composition path, see Table \ref{Table:AtomicComposition}. To select optimal printing parameters, we leveraged previous parameter optimization work for producing fully dense 316L alloy parts on the same L-DED system \cite{Joyce2025Desing}, which followed a well established framework developed by some of the authors \cite{VAUGHAN2023,WHITT2023}. This approach was viable due to the similarities in chemistry, density, and particle size between 316L and the present Fe-Ni-Cr gradient. However, since our gradient was built from elemental powders and not from pre-alloyed material, simultaneous hopper use was necessary, requiring the identification of suitable hopper speed parameters to achieve target chemistries at a given build location. To achieve different target compositions, we calibrated the powder mass flow rate for each powder-hopper combination by collecting flowing powder for 30 seconds at six different hopper speeds (0.25 to 5 RPM) and measuring their weights. This process was repeated three times per powder to minimize uncertainties. The resulting calibration allowed us to determine the required hopper RPMs as a function of target composition for a given total mass flow rate.

The first composition gradient was deposited on an unheated 316L stainless steel substrate using printing parameters used for the 316L stainless steel \cite{Joyce2025Desing}. Specifically, the mass flow rate was set to 0.125~g/s, corresponding to a single hopper operating at 5~RPM. The center purge gas and powder carry gas flows were 20 and 4~l/min, respectively. The laser focal point was positioned 5~mm above the print surface to create a diffuse laser \cite{VAUGHAN2023}, with a laser power of 280~W and a laser scan rate of 8.5~mm/s. Each layer was deposited by first outlining a 10~mm square base, followed by parallel tracks with a 380~\unit{\um} hatch distance, rotated 180\textdegree{} after each pass and offset by 205~\unit{\um} from the contour. After each layer, the beam focal point and printing head were raised by 305~\unit{\um}, with the printing direction rotated 90\textdegree{}. A total of eight layers were deposited per composition section, each building approximately 3~mm of material with a 20\% overbuild. The printing head position was adjusted accordingly to maintain the correct melt pool-to-laser focal point distance. To prevent overheating and maintain printing conditions, the part was allowed to cool for a few minutes after each gradient section. See Table \ref{Table:PrintingParameters} for a summary of the printing parameters.  

Evaluation of the chemical distribution in the first composition gradient revealed that the Cr content deposited during the build was significantly lower than required to achieve the target composition. This discrepancy was likely due to the low Cr mass flow theoretically needed to reach the tenth composition of Ni-3.4~at.\%~Cr, and Cr evaporation during printing. To ensure this composition was achieved throughout the print, a second build was performed using the same methodology but with additional sections incorporating increased Cr mass flow. These added sections guaranteed that the desired Cr content was reached at some point along the gradient, as the Cr concentration increased beyond the originally designed values. The only difference between the designed ''Target'' compositions and the ''Nominal'' compositions used for L-DED was the increased Cr content in these added sections, labeled as a and b in Table 1. This revised build was selected for further characterization and testing, with results presented and discussed in the following sections.

\begin{table}
    \centering
    \begin{tabular}{ccccccc}
        & \multicolumn{3}{c}{Target, at. \%} & \multicolumn{3}{c}{Nominal, at. \%}\\
        Section & Fe & Ni & Cr& Fe & Ni & Cr\\ \hline
        1 & 82.7 & 17.3 & 0 & 82.7 & 17.3 & 0\\
        2 & 80.5 & 19.5 & 0 & 80.5 & 19.5 & 0\\
        3 & 78.2 & 21.8 & 0 & 78.2 & 21.8 & 0\\
        4 & 75.7 & 24.3 & 0 & 75.7 & 24.3 & 0\\
        5 & 72.8 & 27.2 & 0 & 72.8 & 27.2 & 0\\
        6 & 69.6 & 30.4 & 0 & 69.6 & 30.4 & 0\\
        7 & 65.3 & 34.7 & 0 & 65.3 & 34.7 & 0\\
        8 & 58.4 & 41.6 & 0 & 58.4 & 41.6 & 0\\
        9 & 28.9 & 71.1 & 0 & 28.9 & 71.1 & 0\\
        10 & 0 & 96.6 & 3.4 & 0 & 89.2 & 10.8\\
        a & & & & 0 & 85.3 & 14.7\\
        b & & & & 0 & 81.8 & 18.2\\
    \end{tabular}
    \caption{Designed target compositions and nominal compositions at each section of the linear SFE-graded part. Target compositions correspond to those prescribed by the designed gradient, whereas nominal compositions correspond to those calculated from the powder mass flow rates recorded during fabrication via L-DED.}
    \label{Table:AtomicComposition}
\end{table}

\begin{table}
    \centering
    \begin{tabular}{cccccccccc}
         &  &  &  &  &  &  &  &  & \\
         Carry Gas F.R&  4 l$\cdot$min$^{-1}$&  Beam Power&  280 W&  Hatch Distance&  380 \unit{\um} &  Layer Shape&  Square&  & \\
         Purge Gas F.R.&  20 l$\cdot$min$^{-1}$&  Scan Rate&  8.5 mm$\cdot$s$^{-1}$&  Contour Offset&  205 \unit{\um}&  Layer Side Length&  10 mm &  & \\
         Mass F.R.&  0.125 g$\cdot$s$^{-1}$&  Pattern Rotation&  90 deg &  Layer Thickness&  305 \unit{\um}&  Layers per Section&  8&  & \\
    \end{tabular}
    \caption{Summary of L-DED process parameters used for the fabrication of the gradient. F.R. stands for flow rate.  }
    \label{Table:PrintingParameters}
\end{table}

After L-DED, the linear SFE gradient was cut from the substrate and cross-sectioned along the build direction using wire electrical discharge machining (EDM). The two cross-sections were then polished to remove the re-cast layer from EDM and produce a suitable surface finish for characterization. One of the halves was kept in the as-printed condition, and the second half was sealed in a quartz tube under a protective atmosphere of high-purity Ar (99.999\%), heat treated at 800~°C for 1~hour, and quenched using room temperature water. We will refer to these samples as ``As Printed'' and ``Heat Treated'', respectively.

\subsection{{Sample Characterization}}
\subsubsection{Microstructural Characterization}

All polishing and grinding was conducted manually on an Allied Metprep 3 Autopolisher. Polishing was achieved through steps beginning with grinding on 400 grit sandpaper until the sample was flat. This was followed by polishing with diamond suspensions of 9, 3, and 1~\unit{\um} before a finishing step was conducted with 0.05~\unit{\um} colloidal silica. The variety of compositions resulted in an iterative process where the required time per step varied between individual layers. The sample was repeatedly imaged in a Phenom XL desktop to determine when the polish obtained by each step was sufficient to move to the next step. Room temperature X-ray diffraction (XRD) was performed to characterize the microstructure of both As Printed and Heat Treated samples using a Bruker D8 Discover XRD system. This system was equipped with a Cu X-ray source, a collimator with an aperture of 1~mm, and a laser for targeting the desired diffraction area on the sample. This permitted the acquisition of XRD spectra at different sections along the SFE gradient. Specifically, 12 XRD spectra were obtained along the build direction using a step size of 3~mm. Background intensity due to fluorescence, especially caused by Fe, was subtracted.

An FEI Quanta 600 field emission scanning electron microscope (FE-SEM) and a Phenom XL desktop SEM were employed for imaging and chemical analysis of the As Printed and Heat Treated samples. The composition gradient was characterized by performing multiple energy dispersive spectroscopy (EDS) measurements along the build direction.

\subsubsection{Mechanical Property Characterization}

Both microindentation and nanoindentation were employed to map the property changes along the compositional gradient for both the As Printed and Heat Treated samples. To prepare the As Printed sample for indentation, the sample was mounted in a standard 31.8~mm metallographic mount using an Allied High Tech TechPress3 and Allied's graphite-filled, phenolic mounting resin. It should be noted that the As Printed sample was approximately 34.7~mm long, making the sample too large to mount in the available mounting press. Therefore, the sample edges were ground down until the sample was approximately 31.2~mm long. This resulted in $\sim$2.5~mm lost from the beginning of the sample build and $\sim$1.0~mm lost from the end of the sample. In contrast, the Heat Treated sample was mounted directly onto a metal block with Crystalbond 509 to allow metallographic polishing without reducing the overall length of the sample. This difference in sample length has been accounted for in reporting results for both the As Printed and Heat Treated samples.

For nanoindentation on the As Printed sample, a continuous row of 61 nanoindents was made with a prescribed spacing 0.5~mm, starting from a position $\sim$0.5~mm from the end of the mounted sample. Conversely, the Heat Treated sample had a continuous row of 69 nanoindents with the same 0.5~mm spacing and starting position $\sim$0.5~mm from the end of the sample. The different number of indents in the two samples was due to the difference in sample length, as described above. The 0.5~mm spacing was used for almost all indents unless the indent would have landed on a defect, in which case the indent was moved at least 3 indent diameters ($\sim$60~\unit{\um}) away, per the recommendations outlined in ISO 14577-1 \cite{ISO14577-1}. To evaluate the hardness, elastic modulus, and strain rate sensitivity (SRS) of each composition, strain rate jump tests (SRJT) were performed instead of constant strain rate tests. The background and reasoning behind nanoindentation SRJT are explained in-depth by Maier-Kiener and Durst \cite{maierkiener2017advanced}. An initial strain rate, $\dot\varepsilon$, of 0.1~s$^{-1}$ was chosen with jumps to 0.01~s$^{-1}$ and 0.001~s$^{-1}$, where each jump to a lower strain rate is followed by a jump to the initial 0.1~s$^{-1}$ strain rate. It should be noted that the indentation strain rate $\dot\varepsilon = \dot{h}/h$ is controlled with the loading rate (i.e., $\dot{P}/P$) of the indenter. The two are related by equation $\dot\varepsilon\approx \frac{1}{2}(\dot{P}/{P})$ \cite{maierkiener2017advanced, lucas1999indentation}. The jump timing is indentation depth controlled, thus, the test starts at a strain rate ($\dot{h}/h$) of 0.1~s$^{-1}$, then jumps to 0.01~s$^{-1}$ at 1000~nm, jumps back to 0.1~s$^{-1}$ at 1500~nm, jumps down to 0.001~s$^{-1}$ at 2000~nm, and finally jumps back to 0.1~s$^{-1}$ at 2500~nm. Additionally, all nanoindentation tests were performed using continuous stiffness measurements where hardness and elastic modulus values were averaged from 2600~nm to 2900~nm. This depth range ensures a more accurate portrayal of the bulk sample at a typical strain rate for nanoindentation (0.1~s$^{-1}$) while also minimizing potential indentation size effects. Each nanoindent is completed once either a depth of 3000~nm or a load of 1000~mN has been reached.

Microhardness data was collected from both the As Printed and Heat Treated sample using a Vickers indenter according to the ASTM E92-17 standard \cite{ASTME92-17}. For the As Printed sample, data was collected from a row of 40 indents along the compositional gradient. For the Heat Treated sample, data was collected from a row of 54 indents along the compositional gradient. Again, more indents were used on the Heat Treated sample because of its longer length due to the different metallographic preparation techniques described above. Three indents landed on pores on the Heat Treated sample surface and yielded inaccurate data, resulting in a total of 51 data points for that sample. For all sets of indents, the chosen spacing between indents (0.65~mm) was much greater than the minimum ASTM E92-17 \cite{ASTME92-17} required spacing of 2.5$\times$ the maximum Vickers diagonal (0.188~mm in this work). A test force of 300~gf with a 10-second dwell time was used for all Vickers indents.

\section{Results}

\subsection{{Chemical and Microstructural Characterization}}

\begin{figure}
    \centering
    \includegraphics[width=0.75\linewidth]{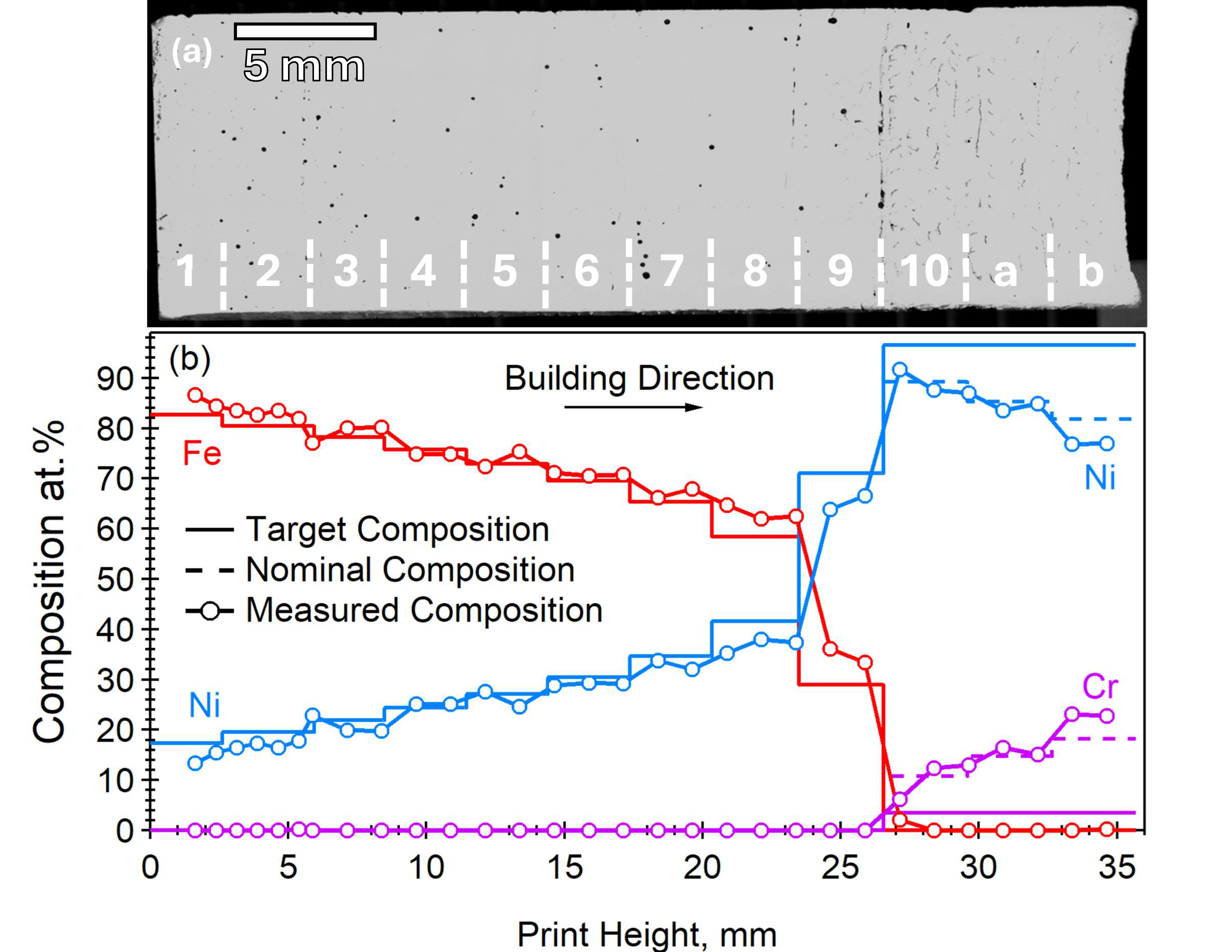}
    \caption{(a) Optical micrograph and (b) the printing height dependence of the target (simple solid lines), nominal (dashed lines) and actual (open symbols) composition as measured by EDS for the As Printed sample. The locations of each composition section are indicated in panel (a).}
    \label{fig:EDSandOM}
\end{figure}

For evaluating the quality of the additively manufactured compositional gradient, an optical image of the cross-section of the Heat Treated sample is presented in Fig. \ref{fig:EDSandOM}a. The build direction extends from left to right.  At this scale, the appearance of the As Printed and Heat Treated samples are essentially the same. The defect distribution within the sample can be categorized into two distinct regions. The first region comprises compositional sections 1–9 (as listed in Table \ref{Table:AtomicComposition}), with measured compositions ranging from Fe-14~at.\%~Ni to Ni-35~at.\%~Fe. The second region includes section 10 and the additional sections a and b, with compositions varying from Ni-6~at.\%~Cr-2~at.~\%Fe to Ni-23~at.\%~Cr. The first region exhibits non-uniformly distributed porosity, while the second region contains fewer pores but features linear voids, indicating possible lack-of-fusion defects. This issue is particularly pronounced at the interface between these two regions, where a large unbonded area perpendicular to the build direction is observed. The observed defect distribution correlates with the evolution of Fe, Ni, and Cr content across the gradient, as shown in Fig. \ref{fig:EDSandOM}b. Notably, the interface between the two regions—and the location of the most significant void—aligns with the transition from Ni-Fe to Ni-Cr compositions.

Fig. \ref{fig:EDSandOM}b compares the target, nominal, and measured compositions of the gradient, corresponding to the determined linear SFE composition path, the expected composition based on mass flow rate calculations, and the composition measured via SEM-EDS, respectively. The target and nominal compositions deviate only for sections 10, a, and b, as shown in Table \ref{Table:AtomicComposition}, where the Cr hopper speed was increased trying to ensure an adequate amount of Cr in the final gradient. The measured composition closely follows the designed chemical gradient path from sections 1 to 9. The largest deviation occurs in section 9, aligning with the most significant change in chemistry. From section 10 to b, the measured composition aligns well with the nominal composition, resulting in higher Cr contents than the originally intended target composition of Ni-3~at.\%~Cr in section 10.

In the initial print, a hopper speed of 0.17 RPM, expected to deliver a Ni-3~at.\%~Cr mixture, resulted in minimal Cr deposition (0.25~at.\%). In contrast, in the second print, hopper speeds of 0.5, 0.75, and 1 RPM resulted in compositions in good agreement with the determined mass flow rate-hopper speed relationship. The under-deposition of Cr was observed only at very low hopper speeds, making it challenging to precisely achieve certain compositions with low Cr content. However, increasing the hopper speed restored the expected system behavior, albeit with some composition overshooting. Consequently, the closest match to the target composition in section 10 was reached early, with a measured composition of Ni-6~at.\%~Cr-2~at.\%~Fe. 

\begin{figure}
    \centering
    \includegraphics[width=0.75\linewidth]{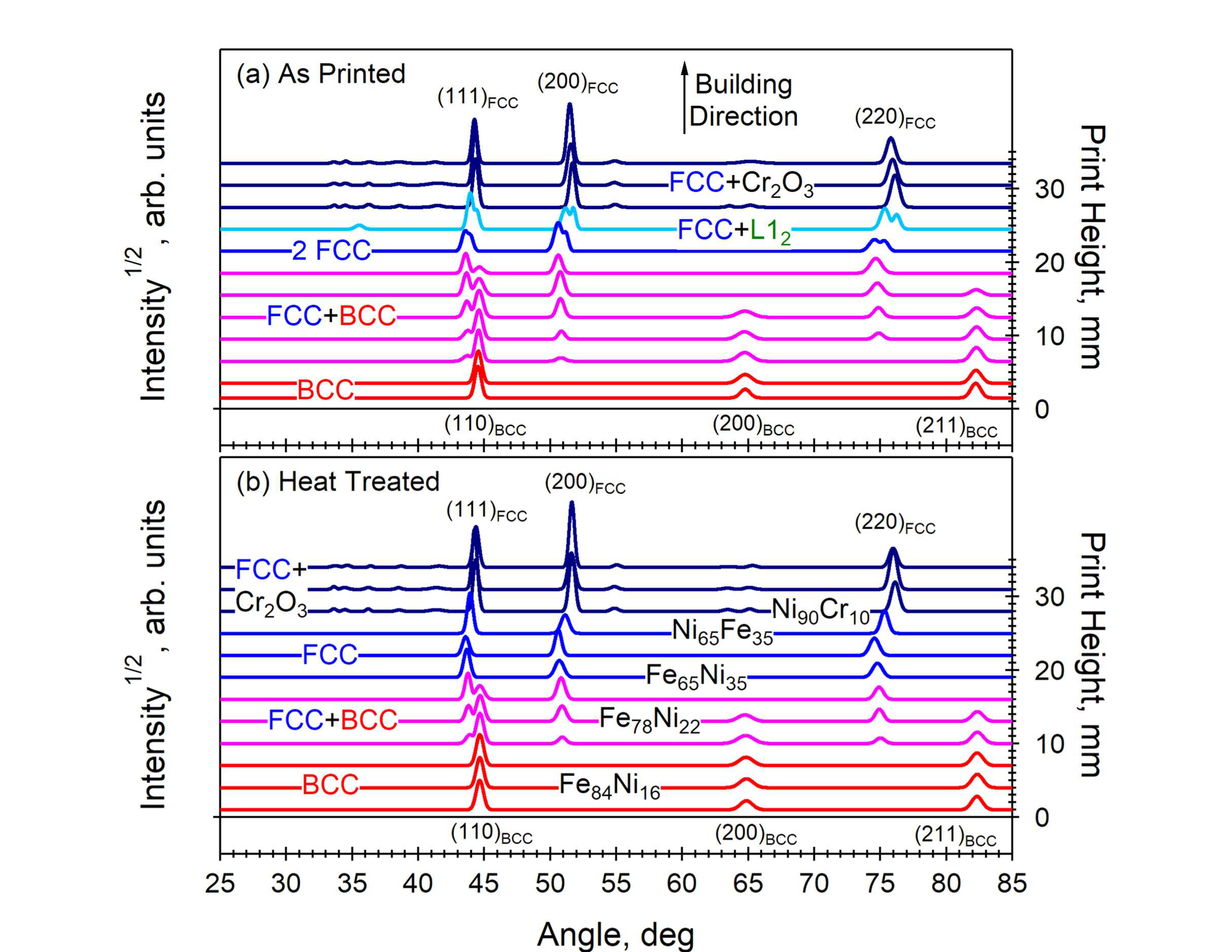}
    \caption{Room temperature XRD spectra obtained at different positions along the build direction for the (a) As Printed and (b) Heat Treated samples. Colors denote areas characterized by the presence of different phases. In (b), some measured chemical compositions are noted for comparison.}
    \label{fig:XRD}
\end{figure}

XRD spectra of both the As Printed and Heat Treated samples were obtained at room temperature at different points along the build direction to characterize the dependence of alloy phases on composition. In Fig. \ref{fig:XRD}a, the As Printed sample presents a strong dependence on composition/SFE, but also presents significant chemical inhomogeneities. Starting from the bottom of the figure, where the gradient is heavily Fe-rich and calculated SFE is close to the very negative endpoint value of -826.2~$\frac{mJ}{m^2}$, the alloy presents a single-phase BCC structure. Upon substituting Fe by Ni and increasing SFE (as predicted by the model), there is a gradual transition from BCC to FCC phases.  However, once the BCC phase has vanished, the FCC phase presents doublets indicating that this section of the gradient is not chemically homogeneous. Furthermore, the spectrum corresponding to section 9 presents a weak peak where the diffraction angle agrees with the expected value for the $(100)_{L1_2}$ peak for one of the FCC phases, suggesting a certain degree of $L1_2$-type atomic ordering, characteristic of the Ni$_3$Fe phase \cite{VANDEEN1981}. Note that section 9 corresponds to an abrupt change in composition, where Ni becomes the majority atomic species and SFE becomes positive. Finally, on the Ni-rich Ni-Cr region, the gradient presents a single FCC phase, but also several weak diffraction peaks suggesting the presence of Cr$_2$O$_3$ impurities introduced during the AM process.

The appearance of these different phases and chemical inhomogeneities was the motivation to homogenize one of the parts at 800~°C for 1 hour followed by water quenching. The room temperature XRD results for this Heat Treated sample are presented in Fig. \ref{fig:XRD}b. These new spectra are missing many of the characteristics seen in the As Printed sample. There were no doublets in the FCC region, nor were there $L1_2$ ordering peaks, demonstrating that the heat treatment was effective in reducing these microstructural features. Notably, the Heat Treated sample still presented a single BCC phase for the most Fe-rich section, and the two-phase region had narrowed as compared to the As Printed sample. Overall, the Heat Treated sample demonstrated a much wider single FCC phase region, although it still presented Cr$_2$O$_3$ impurities in sections 10 to b. Another significant observation is the large shift in the diffraction angle of the FCC peaks between the spectra taken between approximately 22 and 28~mm (sections 8 to 10). This region contains a sharp change in composition, from Fe-rich Fe-Ni at $\sim$22~mm to Ni-rich Fe-Ni at $\sim$25~mm and then to Ni-rich Ni-Cr at $\sim$28~mm. Despite all these compositions presenting a single FCC phase, this chemistry change results in a reduction of $\sim$5.2\% in volume per atom, which equates to an increase of $\sim$7.4\% in density. This large difference in volume may be responsible for the lack of bonding defects observed between sections 9 and 10 in Fig. \ref{fig:EDSandOM}a.

\begin{figure}
    \centering
    \includegraphics[width=0.75\linewidth]{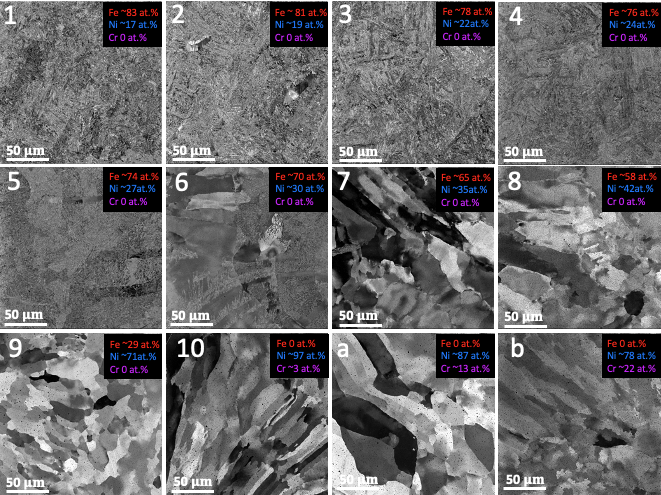}
    \caption{Backscattered electron micrographs from each section of the Heat Treated sample. Labels correspond with the locations denoted in the optical micrograph in Fig \ref{fig:EDSandOM}.}
    \label{fig:SEM}
\end{figure}

Fig. \ref{fig:SEM} presents backscattered electron micrographs obtained for each section of the Heat Treated sample. These micrographs emphasize the transition from BCC phase in sections 1 to 3 to FCC phase in sections 7 to b. The former phase demonstrated fine laths, while the latter was composed of large grains. In addition, the coarser microstructure of the FCC phase allows for the observation of a fine dispersion of oxides, originating from the additive manufacturing process. It is also noted that the phase transition occurred gradually along sections 4-6, matching with XRD data in Fig. \ref{fig:XRD}b. 

\subsection{Mechanical Property Mapping}

\begin{figure}
    \centering
    \includegraphics[width=0.75\linewidth]{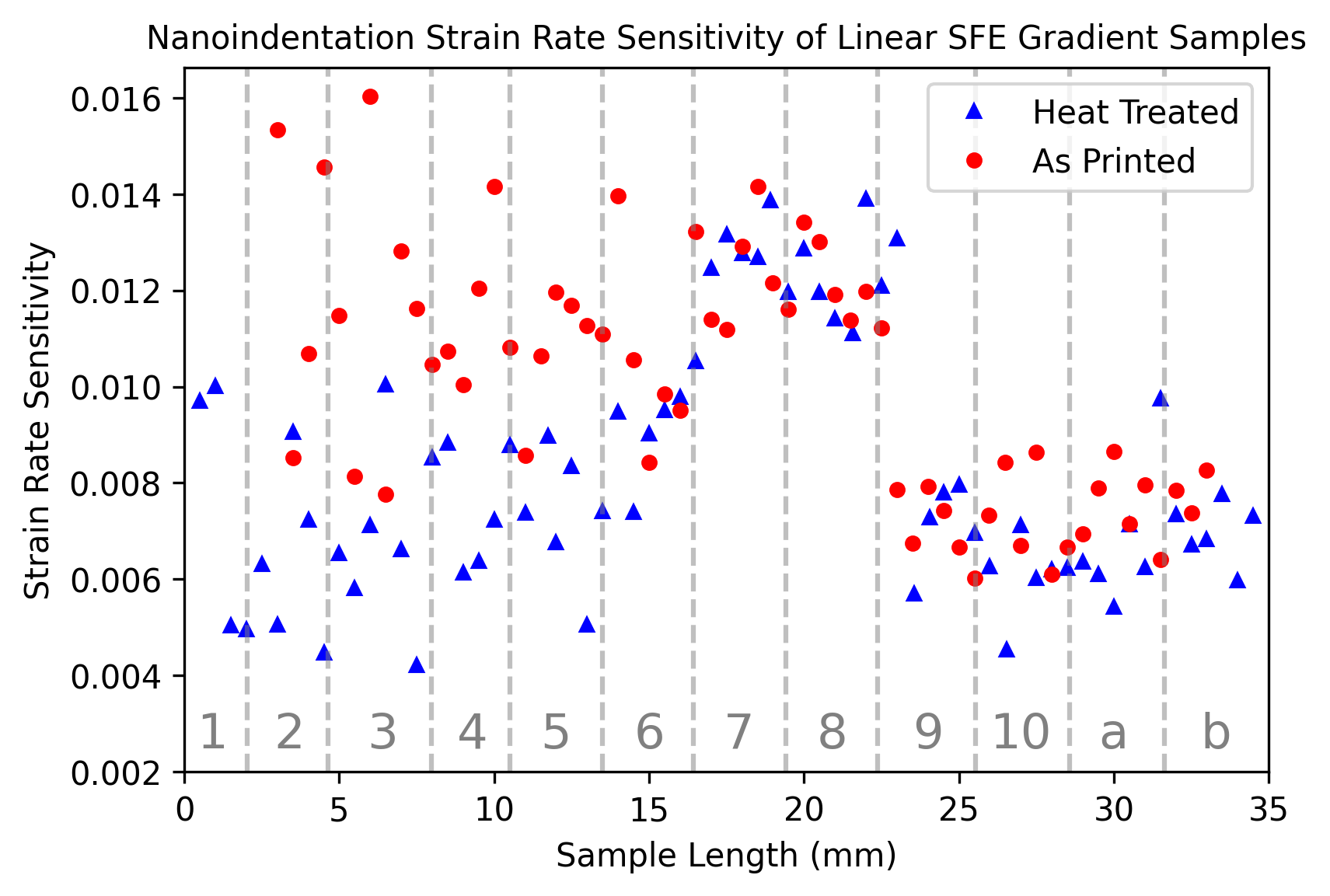}
    \caption{Nanoindentation strain rate sensitivity results for the As Printed and Heat Treated samples. The dotted lines and labels distinguish different compositional sections, which correspond with the locations denoted in the optical microscopy image in Fig \ref{fig:EDSandOM}.}
    \label{fig:NI-SRS-Results}
\end{figure}

\begin{figure}
    \centering
    \includegraphics[width=0.75\linewidth]{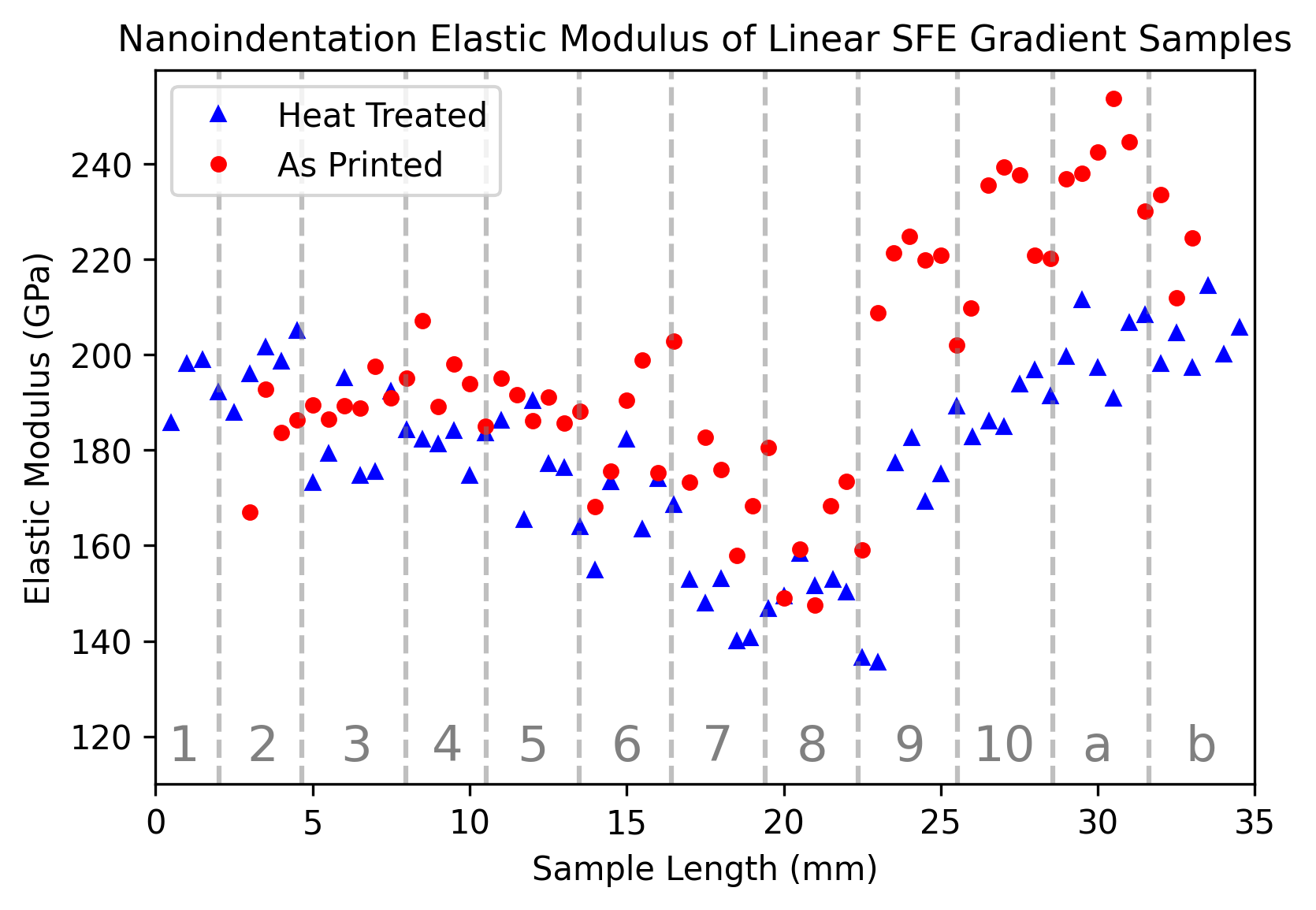}
    \caption{Nanoindentation elastic modulus results for the As Printed and Heat Treated samples. The dotted lines and labels distinguish different compositional sections, which correspond with the locations denoted in the optical microscopy image in Fig \ref{fig:EDSandOM}.}
    \label{fig:NI-E-Results}
\end{figure}

\begin{figure}
    \centering
    \includegraphics[width=0.75\linewidth]{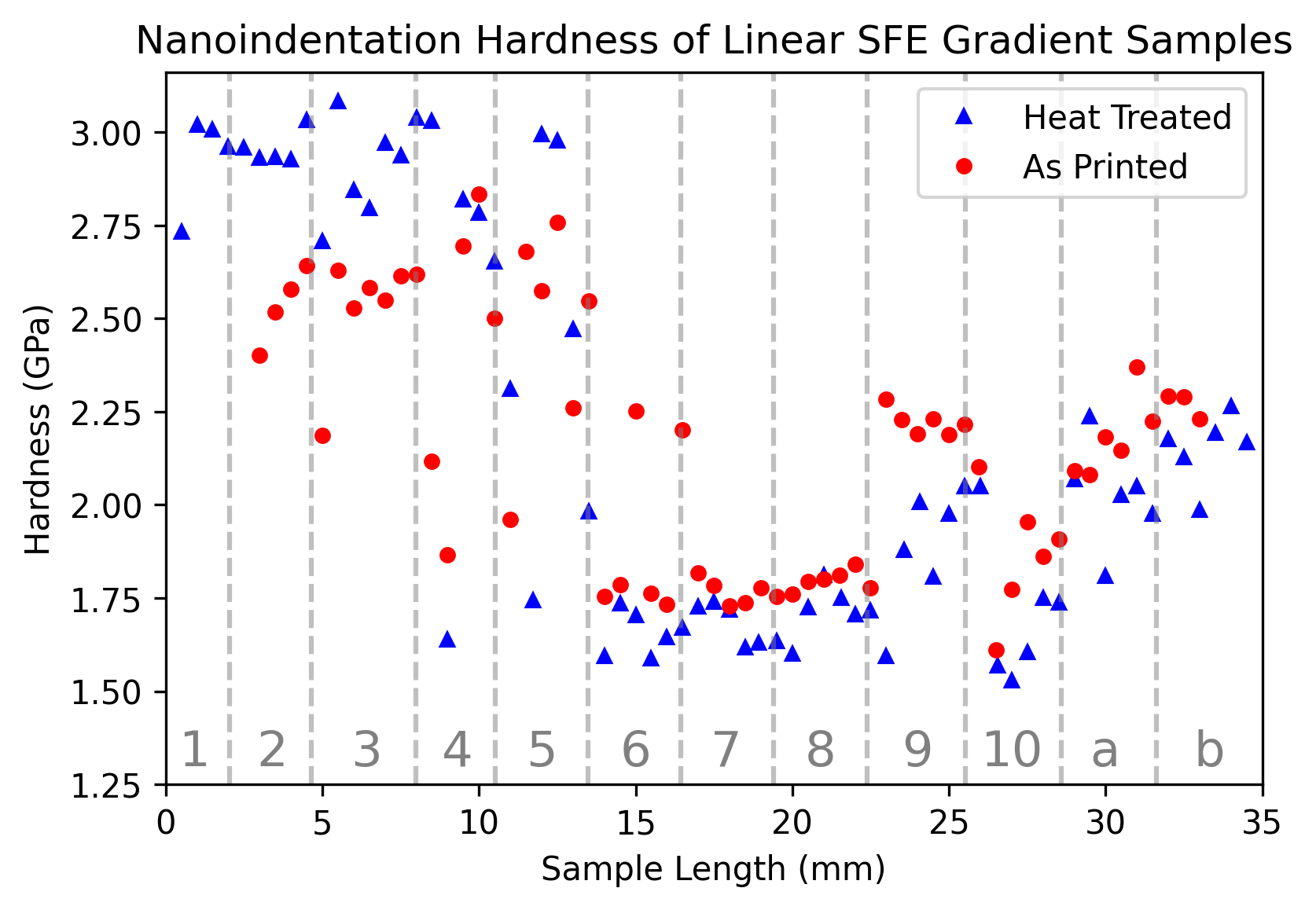}
    \caption{Nanoindentation hardness results for the As Printed and Heat Treated samples. The dotted lines and labels distinguish different compositional sections, which correspond with the locations denoted in the optical microscopy image in Fig \ref{fig:EDSandOM}.}
    \label{fig:NI-H-Results}
\end{figure}

In Figures \ref{fig:NI-SRS-Results}-\ref{fig:NI-H-Results}, the compositional regions have been aligned with nanoindentation results using distinct features from the build (i.e., print layer boundaries) to match the regions shown in Fig. \ref{fig:EDSandOM}. Considering the nanoindentation results as a whole, the As Printed sample exhibits less distinct trends and more scatter in the data, which is most likely due to the presence of unintended complex phases; however, the general trends are still present in both samples. The SRS results (Fig. \ref{fig:NI-SRS-Results}) show a very defined decrease in SRS moving from sections 8 to 9, corresponding to the first large decrease in Fe content (Fig. \ref{fig:EDSandOM}) reducing the composition of Fe to less than 50~at.\% and making Ni the dominant element. There is also a slight increase in SRS between sections 6 and 7 that is more obvious in the Heat Treated data. This increase correlates to the transition from a mixture of BCC and FCC phases to a single FCC phase (Fig. \ref{fig:XRD}). The elastic modulus results (Fig. \ref{fig:NI-E-Results}) show a gradual decrease until approximately section 8, followed by an increase from section 9 onward. This change in elastic modulus seems to mirror the change in composition, with the change from decreasing to increasing elastic modulus paralleling the change in composition from being Fe-dominated to Ni-dominated (Fig. \ref{fig:EDSandOM}). The nanoindentation hardness results (Fig. \ref{fig:NI-H-Results}) display two distinct data trends. Despite scatter in the data, both the As Printed and Heat Treated samples display a drop in hardness occurring between sections 4 and 6. This is consistent with the changing phases from BCC to FCC (Fig. \ref{fig:XRD}) and is also reflected in the SEM images showing a changing microstructure from section 4 to 6 (Fig. \ref{fig:SEM}). The other distinct data trend is the drop in hardness between sections 9 and 10, which corresponds to the complete loss of Fe in the sample (Fig. \ref{fig:EDSandOM}).

\begin{figure}
    \centering
    \includegraphics[width=0.75\linewidth]{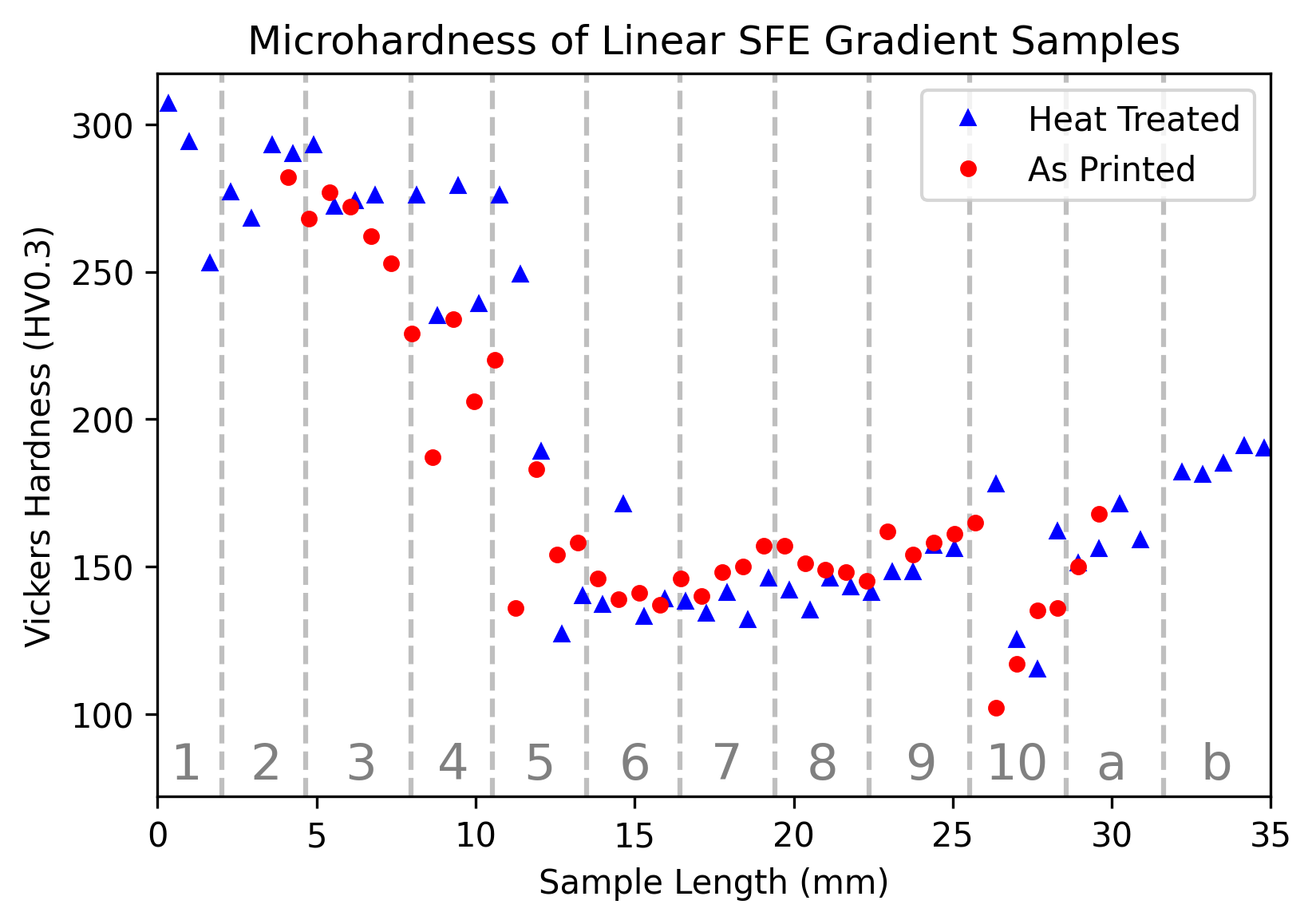}
    \caption{Microhardness results for the As Printed and Heat Treated samples. The dotted lines and labels distinguish different compositional sections which correspond with the locations denoted in the optical microscopy image in Fig \ref{fig:EDSandOM}.}
    \label{fig:MH-Results}
\end{figure}

As mentioned for the nanoindentation results, the microindentation results for the As Printed sample have been shifted to account for sample loss and match the Heat Treated sample. Furthermore, as previously noted, the build direction is from left to right with the beginning of the build at zero. The compositional regions are approximately outlined in Fig. \ref{fig:MH-Results}. Looking at the results, there is a notable drop in hardness starting around section 4 and ending around section 6, with the drop being more defined in the Heat Treated sample. This drop corresponds to the obvious changing microstructure from section 4 to 6 in Fig. \ref{fig:SEM} as well as the changing phases from BCC to FCC as seen in Fig. \ref{fig:XRD}. There is an additional drop in hardness from sections 9 to 10, which occurs for both samples equally and correlates with the loss of Fe going from sections 9 to 10 and onward (Fig. \ref{fig:EDSandOM}). The same general data trends are present in the nanoindentation hardness results, despite the added scatter from the As Printed sample.

\section{Discussion}

\subsection{Microstructural Mapping}

Despite the gradient being designed to contain a single FCC phase above 600~\textdegree{C}, the sample heat treated at 800~\textdegree{C} and quenched did not yield a single FCC phase, but rather a combination of FCC and BCC phases as suggested by XRD. Upon viewing the microstructure in the sample, it is likely that the Fe-rich compositions underwent a martensitic transformation during quenching, and the BCC phase observed by XRD is martensite \cite{zwell1970lattice}. Images 1-4 in Fig \ref{fig:SEM} exhibit an almost completely martensitic microstructure, while 5 and 6 contain some martensite in addition to retained austenite. These images correspond to $\sim$27 and $\sim$30~at.\% Ni, respectively. This suggests that at these concentrations the martensite finish temperature $(M_\mathrm{f})$ was below the temperature of the water used for quenching (i.e. room temperature), while the martensite start temperature $(M_\mathrm{s})$ was above it. At lower concentrations of Ni, both $M_\mathrm{s}$ and $M_\mathrm{f}$ lie above the water temperature, hence a fully martensitic microstructure was achieved. Meanwhile, images 7 to b in Fig \ref{fig:SEM} show a fully austenitic microstructure, meaning that at higher Ni concentrations both $M_\mathrm{s}$ and $M_\mathrm{f}$ lie below room temperature. These results are generally in agreement with $M_\mathrm{s}$ and $M_\mathrm{f}$ temperatures measured for Fe-Ni alloys by Thome et al. using differential scanning calorimetry (DSC) \cite{thome2023local}. They measured that $M_\mathrm{f}$ drops below room temperature in alloys with more than $\sim$26~at.\% Ni.

The Fe-rich end of the gradient was predicted to have an extremely negative SFE by the regression model used for path planning. It is worthwhile to note again that the SFE for this model is being defined as the difference in energy between FCC stacking and HCP or DHCP stacking. An extremely negative SFE does not necessarily have a practical physical meaning and cannot be measured experimentally, but it is indicative of the instability of FCC-like stacking. The martensitic transformation in these composition ranges is thus not surprising. The microstructural results highlight a key limitation of the methods employed in this investigation. That is, that although a set of compositions may be predicted by equilibrium CALPHAD calculations to be completely single-phase at high temperatures, they are not guaranteed to retain that phase upon quenching. Future investigations must take this into account if a single phase is desired, possibly by screening composition paths with information from literature on behavior of selected compositions when quenched. 

\subsection{Mechanical Property Mapping}

There are two clear trends in the hardness results, present in both the nanoindentation (Fig. \ref{fig:NI-H-Results}) and microhardness (Fig. \ref{fig:MH-Results}) plots. The first is a noticeable drop in hardness from sections 4 to 6. This initial hardness drop mirrors the transition from a fully martensitic microstructure to a martensite and retained austenite microstructure. As is well-documented in the literature, adding austenite to a martensitic microstructure is known to lower the hardness \cite{MSENIntroCallister, Tabatabae2011InfluenceofRetainedAustenite, Sherby2008RevisitingtheStructure, Barlow2012EffectofAustenitizing}. The second hardness trend is a drop in hardness from sections 9 to 10, which correlates to a large drop in, and almost complete loss of, Fe. The composition of section 9 was measured as 65~at.\% Ni and 35~at.\% Fe, while the composition of section 10 was measured to be 92~at.\% Ni, 6~at.\% Cr, and 2~at.\% Fe. It is known that Fe (as well as Cr) can be used to harden Ni through solid solution strengthening, which is related to lattice strain due to the differing atomic diameters \cite{ASMHandbook}. Therefore, the almost complete loss of Fe, resulting in a 92~at.\% majority of Ni with only small amounts of Cr and Fe, could have caused the observed drop in hardness. Using this same reasoning, as the amount of Cr increases to $\sim$15~at.\% in section a and further to $\sim$23~at.\% in section b, the resulting increase in hardness would also be explained by solid solution strengthening.

The most noticeable trend in the SRS data (Fig. \ref{fig:NI-SRS-Results}) is a comparatively large drop moving from sections 8 to 9. This drop in SRS does not correlate to any change in existing phases as seen in Fig. \ref{fig:XRD}, but does correlate with Ni overtaking Fe as the dominant element (Fig. \ref{fig:EDSandOM}a) as well as the predicted SFE going from a negative to positive value (Fig. \ref{fig:sfigLin}). The authors are not aware of any current research that explains this phenomenon in relation to the noticed correlations. However, negative SFE seems to imply that the stacking sequence is metastable and the material would prefer to be a different stacking sequence \cite{Zhang2017OrignofNegativeSFE, Shih2021StackingFaultEnergy}. Thus, the material should be more sensitive to strain rate (i.e., higher SRS). Conversely, a low positive SFE means dislocations can more readily separate into partials, requiring screw dislocations to recombine before cross-slip can occur.  Huang et al. has shown that low-SFE materials retain mobile dislocations longer during deformation, resulting in a lower SRS \cite{Huang2014StrainRateSensitivity}.

The main trend in the elastic modulus measurements from nanoindentation (Fig. \ref{fig:NI-E-Results}) is a slightly decreasing elastic modulus from sections 1 to 8 and then an increasing elastic modulus from sections 9 to 10 and onward, with the lowest elastic modulus values between sections 8 and 9. This data trend is reflected in the literature and most obvious in the compilation report by Ledbetter and Reed, which comprehensively looks at the elastic properties of Fe-Ni alloys \cite{LedbetterReed1973Elastic}. They found that the elastic modulus of Fe-Ni alloys shows a minimum at $\sim$40~wt.\% Ni ($\sim$39~at.\% Ni), which agrees with the elastic modulus minimum in Fig. \ref{fig:NI-E-Results} that corresponds to an actual composition of $\sim$38~at.\% Ni (Fig. \ref{fig:EDSandOM}). Ledbetter and Reed noted that this trend could be related to the differing atomic sizes of solute and solvent atoms, resulting in residual strain energy.

Another reason for the reduction in elastic modulus in Fe-Ni alloys with Ni contents around 35-40 at.\% as compared to those with lower Ni contents can be due to the elastic softening phenomenon occurring prior to martensitic transformation. This is observed in several martensitically transforming alloys including several $\beta$-Ti alloys such as TiNb and many shape memory alloys such as NiTi, NiTiCu, NiTiFe, CoNiAl, CuNiAl, NiMnGa, etc. \cite{tane2011low, ren2001comparative, ren1999elastic, ren1999understanding, jeong2010elastic, recarte2004study, heczko2018temperature, seiner2013combined}. As the material approaches the martensitic transformation temperature upon cooling, all these martensitically transforming materials exhibit softening of shear modulus c' or both c' and c$_{44}$ and low measured elastic modulus values. In the current Fe-Ni materials, since Ni content decreases the martensitic transformation temperature, the elastic modulus at room temperature in the austenite phase drops when the martensitic transformation temperature is below and close to the room temperature.

\section{Summary and Conclusions}
In this work, a functionally graded material (FGM) with a linear gradient in stacking fault energy (SFE) was designed autonomously using a path planning algorithm with endpoint compositions of 82.7~at.\%~Fe, 17.3~at.\%~Ni and 96.6~at.\%~Ni, 3.4~at.\%~Cr. Samples with these compositional gradients were manufactured using L-DED. The primary conclusions that can be drawn from the analysis of this gradient are as follows:

\begin{itemize}
    \item The composition gradient design framework based on the RRT*FN path planning algorithm has been proven to be a viable method for application to high-throughput alloy design, synthesis, and characterization workflows. Because of its ability to plan monotonic property paths, this framework would be particularly useful for investigating correlations between design criteria and properties of interest such as hardness.
    \item Although the entire FGM was predicted to be single-phase FCC at high temperatures by CALPHAD, it did not retain that characteristic upon heat treatment and quenching. Instead, much of the Fe-rich end of the gradient underwent a martensitic transformation upon cooling. This demonstrates another potential application of this FGM design framework: determining where limits exist for various models used in alloy design. In this case, equilibrium phase fractions from CALPHAD were not sufficient to ensure a single-phase FGM.
    \item The gradient was planned from the absolute minimum to the absolute maximum predicted SFE among constrained compositions. The compositions with extremely negative SFE values in the gradient corresponded to those that underwent martensitic transformations, highlighting a limit for the SFE model that was built for FCC alloys. However, the negative SFEs do indicate extremely unstable FCC stacking, and therefore help explain why martensitic transformations took place.
    \item Although heat treating the sample altered the phase makeup of the FGM according to XRD results, it did not substantially alter the measured hardness of each composition. The hardest compositions were at the Fe-rich end of the gradient, where the microstructure was almost completely martensitic in the heat treated sample.
    \item The most notable changes in mechanical properties along the gradient corresponded with the transition from a fully martensitic to fully austenitic microstructure, the change from negative to positive SFE, changes in composition from majority Fe to majority Ni, as well as a change from Fe-Ni to Ni-Cr binary alloy compositions.
\end{itemize}

For future work in this area, a number of improvements could be implemented. First, this gradient focused on a far larger range of SFE than is practical to connect to material behavior or possible to measure experimentally. Future investigations should focus only on slightly negative and positive values of SFE to measure and validate the model used. This would additionally allow for testing more compositions in the positive range where experimental measurements and connections to behavior can be established. This improvement would have the additional benefit of removing compositions that exhibited martensitic transformations upon quenching, as they were the same compositions that had extremely negative model-predicted SFE. Another improvement would be the introduction of additional constraints on the gradient. To isolate the effect of SFE on material behavior and mechanical properties, keeping other material properties as constant as is practical would be beneficial. This would involve an additional constraint on the cost function for the RRT*FN algorithm that penalizes deviation in any selected property along the path from a defined mean value. Such properties might include elastic modulus, density, and strength. Ideally, the penalty for any given composition on the path would be proportional to the amount of deviation from the defined property value, similar to how lack of monotonicity in SFE was penalized on the gradient path in this work.

\section*{CRediT authorship contribution statement}
\textbf{James Hanagan:} Writing - Original Draft, Methodology, Software, Visualization, Formal analysis \textbf{Nicole Person:} Writing - Original Draft, Investigation, Visualization, Formal analysis \textbf{Daniel Salas:} Writing - Original Draft, Investigation, Visualization, Formal analysis \textbf{Marshall Allen:} Writing - Original Draft, Methodology, Software, Visualization, Formal analysis \textbf{Wenle Xu:} Visualization, Investigation \textbf{Daniel Lewis:} Investigation, Formal Analysis \textbf{Brady Butler:} Conceptualization, Writing - Review \& Editing, Supervision \textbf{James D. Paramore} Writing - Review \& Editing, Supervision \textbf{George Pharr:} Writing - Review \& Editing, Supervision \textbf{Ibrahim Karaman:} Conceptualization, Writing - Review \& Editing, Supervision, Project administration \textbf{Raymundo Arr\'{o}yave:} Conceptualization, Writing - Review \& Editing, Supervision, Project administration, Funding acquisition

\section*{Data Availability}
The data that supports the findings of this study are available from the corresponding author upon reasonable request.

\section*{Declaration of Competing Interest}
The authors declare that they have no known competing financial interests or personal relationships that could have appeared to influence the work reported in this paper.

\section*{Acknowledgements}
The authors acknowledge the support from DEVCOM-ARL under Contract No. W911NF-22-2-0106 (\emph{BIRDSHOT} Center funded through the High-throughput Materials Discovery for Extreme Conditions (HTMDEC)). The views and conclusions contained in this document are those of the authors and should not be interpreted as representing the official policies, either expressed or implied, of the Army Research Laboratory or the U.S. Government. The U.S. Government is authorized to reproduce and distribute reprints for Government Purposes notwithstanding any copyright notation herein.  Portions of this research were conducted with the advanced computing resources provided by Texas A\&M High Performance Research Computing. We acknowledge the contributions of Michael Elverud during sample fabrication, of Dr. Anup Bandyopadhyay during XRD characterization, and of Matthew Skokan during specimen preparation.

\bibliography{mybibfile}

\end{document}